\documentclass[twocolumn]{IEEEtran}

\usepackage{xcolor,soul,framed} 

\colorlet{shadecolor}{yellow}
\usepackage[pdftex]{graphicx}
\graphicspath{{../pdf/}{../jpeg/}}
\DeclareGraphicsExtensions{.pdf,.jpeg,.png}

\usepackage[cmex10]{amsmath}
\usepackage{amsfonts} 
\usepackage{amssymb}
\usepackage{array}
\usepackage{mdwmath}
\usepackage{mdwtab}
\usepackage{eqparbox}
\usepackage{url}
\usepackage{comment}

\UseRawInputEncoding
\usepackage{amsmath,epsfig,amssymb,verbatim,amsopn,cite,multirow}
\usepackage{amsthm}
\usepackage{balance}
\usepackage{multirow}
\usepackage{amsfonts}
\usepackage{amssymb}
\usepackage{epsfig}
\usepackage{epstopdf}
\usepackage{bm}
\usepackage{balance}
\usepackage{graphicx}
\usepackage{silence}
\usepackage{subcaption}
\usepackage{footnote}
\usepackage[margin=15mm,top=18mm,bottom=26mm]{geometry}
  {\proof}{\proofend}
\newtheorem{proposition}{Proposition}

    \title{Cell-Free massive MIMO for ISAC under multi-functional jamming attack}

  \author{Zonghan~Wang,
      Zahra Mobini,
      Hien Quoc Ngo,
     and Michalis Matthaiou
}

\title{\fontsize{0.83cm}{1cm}\selectfont Anti-Malicious ISAC Using Proactive Monitoring}
\author{\IEEEauthorblockN{{Zonghan Wang, Zahra Mobini, Hien Quoc Ngo, and Michail Matthaiou}\\
\IEEEauthorblockA{
Centre for Wireless Innovation (CWI), Queen's University Belfast, U.K.\\
Email:\{zwang95, zahra.mobini, hien.ngo, m.matthaiou\}@qub.ac.uk}}}

\begin{document}


\maketitle

\begin{abstract} 
In this paper, we investigate proactive monitoring  to mitigate malicious activities in integrated sensing and communication (ISAC) systems. Our focus is on a scenario where a cell-free massive multiple-input multiple-output (CF-mMIMO) architecture is exploited by malicious actors. Malicious actors use multiple access points (APs) to illegally sense a legitimate target while communicating with  users (UEs), one of which is suspected of illegal activities.
In our approach, a proactive monitor overhears the suspicious UE and simultaneously sends a jamming signal to degrade the  communication links between the APs and suspicious UE. Simultaneously, the monitor sends a precoded jamming signal toward the legitimate target to hinder the malicious sensing attempts. We derive closed-form expressions for the sensing signal-to-interference-noise ratio (SINR), as well as the received SINR at the UEs and overheard SINR at the monitor. The simulation results show that our anti-malicious  CF-mMIMO ISAC strategy can significantly reduce the sensing performance while offering excellent monitoring performance.
		
\let\thefootnote\relax\footnotetext{
This work was supported by the U.K. Engineering and Physical Sciences Research Council (EPSRC) (grants No. EP/X04047X/1 and EP/X040569/1)
The work of Z. Mobini and H. Q. Ngo was supported by the U.K. Research and Innovation Future Leaders Fellowships under Grant MR/X010635/1. The work of M. Matthaiou was
supported by the European Research Council
(ERC) under the European Union’s Horizon 2020 research
and innovation programme (grant agreement No. 101001331).}
\end{abstract}


%
\IEEEpeerreviewmaketitle


\vspace{-3mm}
\section{Introduction}
ISAC is expected to play a crucial role in next-generation wireless systems. By utilizing infrastructure and resources for both communication and sensing in a cooperative fashion, this powerful technological paradigm aims to enhance the performance of both communicating and sensing functionalities~\cite{ref16,ref25}. 
Recently, the concept of CF-mMIMO has been incorporated into ISAC, in which a large number of communication APs (C-APs) coherently serve multiple UEs  while another set of APs, named as sensing APs (S-APs), are used for sensing operation~\cite{ref9}. CF-mMIMO ISAC architectures with multiple communication and radar devices have triggered significant research interest~\cite{ref:CYu_cellfree_ISAC,10684238}. 

With the current advances in ISAC systems, it is likely that in the near future, we will  witness the extensive ISAC utilization across a spectrum of applications, spanning from  civilian domain, such as smart manufacturing,  vehicular networks, and human activity recognition, to various military operations~\cite{Liu:2022:survey}.
However, if ISAC  technology falls in the hands of adversaries, it can be used as an unauthorized sensing technology, while  unauthorized wireless communication links could also be established to commit (cyber)-crimes or perform illegal activities. These threats highlight the need of adapting security methods from the domains of physical-layer security (PLS) that can ensure public safety{~\cite{ref:yassen_TWC}. }
Surprisingly, the current literature has mainly overlooked the security concerns of ISAC. One crucial security aspect is the protection of a target's location  from being sensed by malicious ISAC as well as the monitoring and interception of suspicious communication links. 

To this end, we are inspired by the emerging PLS technique of proactive monitoring, to propose a new anti-malicious ISAC strategy.
Recently, \textit{proactive monitoring} has started to attract attention  for efficiently monitoring unauthorized communication  links~\cite{ref:zahra_iot}. Specifically,  in proactive monitoring, a legitimate monitor operates in a full-duplex (FD) mode and purposely sends jamming signals to interfere with the suspicious links in order to degrade  the achievable data rate at the suspicious receivers~\cite{ref20,ZAHRA:TIFS:2019}.
The main contributions of this paper are:
\begin{itemize}
    \item We propose an anti-malicious CF-mMIMO ISAC design that utilizes proactive monitoring. In this design, the malicious ISAC system comprises multiple C-APs serving multiple UEs, with one UE suspected of engaging in illegal activities, and multiple S-APs attempting to illicitly sense a legitimate target. In  our anti-malicious  design, the monitor has dual functionalities: it intercepts the transmissions of the suspicious UE and emits a jamming signal to disrupt the communication links between the APs and the suspicious UE. Concurrently, the monitor generates a precoded jamming signal directed at the legitimate target, thereby reducing the probability of successful sensing by the malicious ISAC system. 
    
    \item We provide a detailed theoretic performance analysis of the proposed anti-malicious CF-mMIMO ISAC design and derive closed-form expressions for the SINR  at the  UEs, S-APs, and proactive monitor.  These closed-form expressions facilitate the subsequent system optimization and can shed useful insights into the system performance.
    \item Numerical results show that our anti-malicious CF-mMIMO ISAC design can reduce the sensing detection probability (SDP) of the malicious CF-mMIMO ISAC system, while providing a monitoring successful probability (MSP) around one.
\end{itemize}
\emph{Notation}: We use lower and upper case letters to denote vectors and matrices. The superscripts $(\cdot)^{H}$, $(\cdot)^{\ast}$ and $(\cdot)^{T}$ stand for the Hermitian, conjugate and transpose operators; $||\cdot||$ denotes the Euclidean norm; $\mathbf{I}_N$ stands for the $N \times N$ identity matrix. A zero mean circular symmetric complex Gaussian distribution with variance $\sigma^2$ is denoted by $\mathcal{CN} (0,\sigma^2)$. Moreover, $\mathbb{E}\left \{ \cdot \right \}$ denotes the statistical expectation, while $\mathrm{Var}(\cdot)$ and $\mathrm{Tr}(\cdot)$ denote the variance and trace of a matrix. 
\begin{figure}[t]
  \centering
  \includegraphics[width=0.78\linewidth]{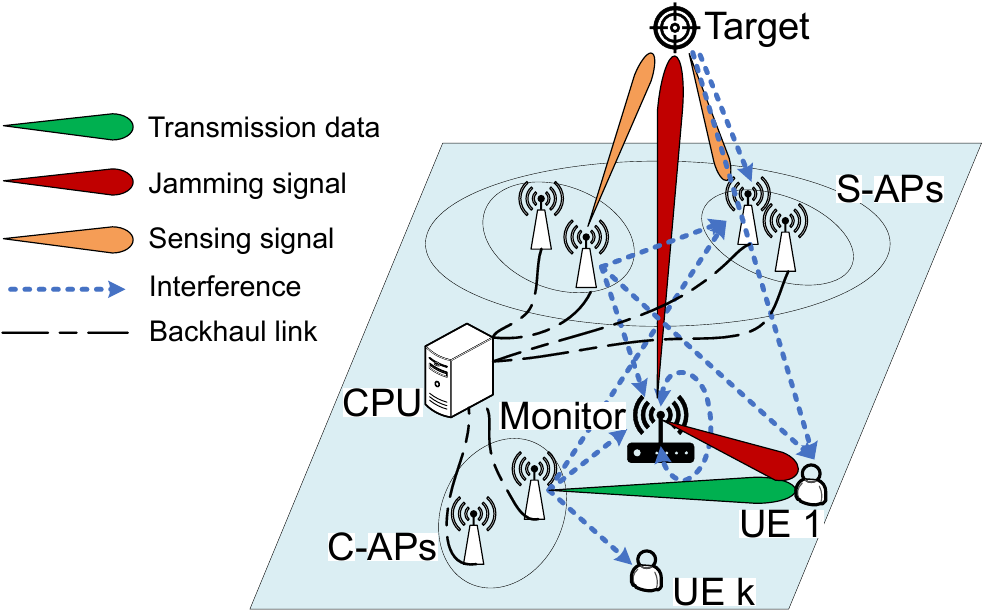}
  \caption{Anti-malicious CF-mMIMO ISAC design using an FD proactive monitor.}
  \label{fig:system model}
\end{figure}
\setlength{\floatsep}{5pt minus 2pt}
\setlength{\textfloatsep}{5pt minus 2pt}
\setlength{\intextsep}{5pt minus 2pt}

\section{System  Model}
In this paper, we consider a malicious ISAC system including $M$ untrusted APs and $K$ untrusted UEs, where the APs are divided into two disjoint sets; i) C-AP set, $ \mathcal{M}_{c}$, for transmitting  signals to  UEs, and ii) S-AP set, $ \mathcal{M}_{s}$, for sensing the location of a legitimate target to determine whether there is a  target or not, where $\mathcal{M}_{c} \cap \mathcal{M}_{s}=\emptyset$. Furthermore, the S-AP set is divided into 2 subsets: i) $\mathcal{M}_{s,t}$   includes the APs serving as  malicious sensing transmitters to transmit probing signals and ii) $\mathcal{M}_{s,r}$ includes the APs serving as sensing receivers to receive echos, where $\mathcal{M}_{s}=\mathcal{M}_{s,t} \cup  \mathcal{M}_{s,r}$ and $\mathcal{M}_{s,t} \cap \mathcal{M}_{s,r}=\emptyset$, as shown in Fig.~\ref{fig:system model}. Against this adversary setup, we develop an anti-malicious ISAC  system comprising an FD monitor, which is deployed to simultaneously: i)  monitor suspicious UE. Without loss of generality, we assume that among
$K$ UEs,   UE $1$ is  suspected
of engaging in illegal activities; 
and ii)  protect the legitimate target from being detected by the malicious ISAC system by jamming the signal towards the target. To be more general, we assume an aerial legitimate target located in 3D space with height $h$ m above the ground. 

We assume that each UE is equipped with a single antenna, while each AP is equipped with $N$ antennas, and the proactive monitor is equipped with $ N_{\mathtt{pm}}$ antennas.

\begin{itemize}
    \item 
The ground-to-ground channel between the $m$-th C-AP ($m \in\mathcal{M}_c$) and the $k$-th UE is modeled as $\mathbf{g}_{m,k}=\beta_{m,k} ^{1/2}\mathbf {g}_{m,k}' $, where $\mathbf{g}_{m,k}'\in \mathbb{C} ^{N\times 1}$ is the small-scale fading vector whose entries are independent and identically distributed (i.i.d.)  $\mathcal{CN}(0,1)$. The value of $\beta_{m,k}$ is the large-scale fading coefficient. The channel between the monitor and UE $k$, $\mathbf{g}_{\mathtt{pm},k}$, can be defined similarly with appropriate modifications.

\item Moreover, it is reasonable to assume that the ground-to-air (air-to-ground) channels are pure line-of-sight (LoS)\cite{ref16}. In particular, the channel between the $m'$-th S-AP, $m'\in\mathcal{M}_{s,t}$ and the target, $\mathbf{h}_{m',t}$, can be written as
\vspace{-1mm}
\begin{equation}~\label{eq:gta_channel}
    \mathbf{h}_{m',t}=\sqrt{\zeta_{m',t} }\boldsymbol{\alpha } _{t}\left ( \phi_{m',t}^{a}, \phi_{m',t}^{e}\right ) ,
\end{equation}
where $\zeta_{m',t} = (\frac{\lambda}{4\pi d_{m',t}} )^{L}$ is the free-space path loss,  ${L}$ is the path loss exponent, $\lambda$ is the wavelength and $d_{m',t}$ is the distance between the $m'$-th AP $(x_{m'},y_{m'},0)$ and the target $(x_{t},y_{t},h)$ in a 3D Euclidean space, which can be given by $d_{m',t}=\sqrt{(x_{m'}-x_{t})^2+(y_{m'}-y_{t})^2+h^2} $.
Moreover, $\boldsymbol{\alpha } _{t}\left ( \phi_{m',t}^{a}, \phi_{m',t}^{e}\right )$ is the steering vector, where $ \phi_{m',t}^{a}$ and $ \phi_{m',t}^{e}$ denote the azimuth and elevation angle of  departure (AoD) from the $m'$-th AP to the target, respectively \cite{ref18}. The same steps can be followed to model the channel between the target and $m''$-th S-AP, $m''\in\mathcal{M}_{s,r}$, denoted by $\mathbf{h}_{t,m''}$.

\item The ground-to-air channel between the target and monitor, $\mathbf{h}_{\mathtt{pm},t}\in \mathbb{C} ^{ N_{\mathtt{pm}} \times 1}$,   can be modeled using~\eqref{eq:gta_channel} with proper changes. 
\item The   air-to-ground channel between the target and UE $k$, $h_{t,k}$, can be modeled using~\eqref{eq:gta_channel} with proper changes.
\end{itemize}
In  the uplink training phase of the malicious ISAC system, the proactive monitor launches a pilot spoofing attack, i.e., sends the same pilot as  UE $1$ to the APs,  to enhance its overhearing performance. 
Following\cite{ref12},  by invoking the minimum-mean-square-error (MMSE)
estimation technique and the assumption of orthogonal pilot
sequences, the estimate of ${\mathbf{g} }_{m,k}$  can be written as $\hat{\mathbf{g} }_{m,k}\sim \mathcal{CN} (\mathbf{0},\gamma_{m,k}\mathbf{I}_N)$,  where 
\begin{align}\label{gamma_mk}
        \gamma_{m,k}
        =\left\{\begin{matrix}
\frac{ \tau _{p}\rho _{p} \beta _{m,1}^{2}}{\tau _{p}\rho _{p} \beta _{m,1}+\tau _{p}\rho _{p,\mathtt{pm}} \beta _{m,\mathtt{pm}}+1}, & k=1 \\
\frac{ \tau _{p}\rho _{p} \beta_{m,k}^{2}}{\tau _{p}\rho _{p} \beta_{m,k}+1}, &k\ne 1,
\end{matrix}\right.
\end{align}
where $\tau_{p}$ denotes the length of the uplink training phase, $\rho_{p}$ and $\rho_{p,\mathtt{pm}}$  are the transmit signal-to-noise ratios (SNRs) for pilot transmission at the UEs and monitor, respectively. 
Let $s_{k}$ be the symbol intended for UE $k$. Then, the signal transmitted by $m$-th C-AP can be written as
\vspace{-1mm}
\begin{align}
\mathbf {x}_{m}=\sum\nolimits_{k=1}^{K}\sqrt{\eta _{m,k}\rho _{\mathtt{c}}} \mathbf {w} _{m,k}^{\mathtt{c}}s_{k},
\end{align}
where $\mathbf {w} _{m,k}^{\mathtt{c}}\in \mathbb{C} ^{N\times 1}$ is the precoding vector generated by the $m$-th C-AP to UE $k$, $\eta_{m,k}$ is the power control coefficient chosen to satisfy the power constraint at the C-APs, and  $\rho_c$ is the normalized downlink SNR. The probing signal sent by the $m'$-th S-AP to the target is given by 
\vspace{-1mm}
\begin{align}
    \mathbf {x}_{m',t}= \sqrt{\eta _{m'\!,t}\rho _{\mathtt{s}}} \mathbf{w}_{m',t}^{\mathtt{s}}s_{t},
    \vspace{-1mm}
\end{align}
where $\mathbf{w}_{m',t}^{\mathtt{s}}\in \mathbb{C} ^{N\times 1}$ denotes the precoding vector for sensing. Moreover, $\eta_{m',t}$ and  $\rho_s$ are the power control coefficient and normalized downlink SNR of the sensing signal, respectively. Additionally, $s_{t}$ is the sensing symbol.

We assume that the monitor has perfect CSI  knowledge. This assumption is reasonable since the monitor can estimate the channels to users via the pilot signals received during the training phase of the malicious ISAC system\cite{ref:perfect_CSI_at_monitor}. In the downlink communication phase, the monitor   sends jamming signals   to disrupt the sensing performance of the malicious ISAC system and to interfere with the transmission links to  UE $1$ using a conjugate beamforming approach.  
The transmitted signal at the monitor is  
\vspace{-1mm}
\begin{align}
    \mathbf {x}_{\mathtt{pm}}= \sqrt{\eta _{\mathtt{pm},t}\rho _{\mathtt{pm}}} \mathbf{w}_{\mathtt{pm},t}^{\mathtt{s}}s_{\mathtt{pm},t}+\sqrt{\eta _{\mathtt{pm},1}\rho _{\mathtt{pm}}} \mathbf{w}_{\mathtt{pm},1}^{\mathtt{c}}s_{\mathtt{pm},1},
    \vspace{-1mm}
\end{align}
where
$s_{\mathtt{pm},t}$ and $s_{\mathtt{pm},1}$ denote the transmit  jamming signal to the target and  UE $1$, respectively. The precoding vectors constructed at the monitor  for the target and   UE $1$  can be given by $\mathbf{w}_{\mathtt{pm},t} ^{\mathtt{s}}=  \mathbf{h}_{\mathtt{pm},t}^{\ast }$ and $\mathbf{w}_{\mathtt{pm},1} ^{\mathtt{c}}=  \mathbf{g}_{\mathtt{pm},1} ^{\ast }$, respectively. Moreover, we consider the conjugate scheme for precoding the probing signal at the $m'$-th S-AP to the target, $m' \in \mathcal{M}_{s,t}$, and at the $m$-th C-AP to UE $k$, $m \in \mathcal{M}_c$, such that $\mathbf{w}_{m',t} ^{\mathtt{s}}=  \mathbf{h}_{m',t} ^{\ast}$ and $\mathbf{w}_{m,k} ^{\mathtt{c}}=  \hat{\mathbf{g}}_{m,k} ^{\ast }$, respectively.
Accordingly, the received signal at the $k$-th UE can be represented as 
\vspace{-1mm}
\begin{align}
\label{eq:y_k_xmk}
    &y_{k}  =\sum\nolimits_{m \in\mathcal{M}_c } \mathbf{g}_{m,k}^{T}\mathbf {x}_{m} \notag\\
       &+\sum\nolimits_{m'\in\mathcal{M}_{s,t} } (\mathbf{g}_{m',k}^{T}+\sqrt{\alpha}h_{t,k}\mathbf{h}_{m',t}^{T})\mathbf {x}_{m',t} \notag\\
       &+(\mathbf{g}_{\mathtt{pm},k}^{T}+\sqrt{\alpha}h_{t,k}\mathbf{h}_{\mathtt{pm},t}^{T})\mathbf {x}_{\mathtt{pm}} +n_{k},
\end{align}
where
  $n_{k}$ represents an additive white Gaussian noise (AWGN) with $n_{k}\sim \mathcal{CN}(0,1) $. Also, $\alpha$ is the target reflection gain that depends on the transmission and reflection coefficient, center frequency and radar cross-section (RCS) of the target \cite{ref16}, calculated by the formula $\alpha = 4\pi \sigma_{\mathtt{RCS}} / \lambda^2$, where $\sigma_{\mathtt{RCS}}$ denotes the RCS of the target.
Moreover, the received signal at the monitor  can be written as
\vspace{-1mm}
\begin{align}\label{eq:yj_xmk}
        &\mathbf{y}_{\mathtt{pm}}= \sum\nolimits_{m \in\mathcal{M}_c }\mathbf{G}_{m,\mathtt{pm}}^{T}\mathbf {x} _{m} \notag \\
        &+ \sum\nolimits_{m'\in\mathcal{M}_{s,t} } (\mathbf{G}_{m',\mathtt{pm}}^{T}+\sqrt{\alpha}\mathbf{h}_{t,\mathtt{pm}}\mathbf{h}_{m',t}^{T})\mathbf{x}_{m',t}  \notag \\
        &+(\mathbf{G}_{\mathtt{pm},\mathtt{pm}}^{T}+\sqrt{\alpha}\mathbf{h}_{t,\mathtt{pm}}\mathbf{h}_{\mathtt{pm},t}^{T})\mathbf {x}_{\mathtt{pm}}+\boldsymbol{n}_{\mathtt{pm}},
\end{align}
where $\mathbf{G}_{\mathtt{pm},\mathtt{pm}} \in \mathbb{C} ^{ N_{\mathtt{pm}} \times N_{\mathtt{pm}}}$ is the self-interference channel between the transmit and receive antennas at the FD monitor which can be modeled through the Rayleigh fading model, and whose elements are i.i.d. $\mathcal{CN} (0,\sigma_{\mathtt{SI}}^2)$ random variables, while $\mathbf{G}_{m,\mathtt{pm}}\in \mathbb{C} ^{N \times N_{\mathtt{pm}}}$ denotes the channel between the $m$-th AP and the monitor. Moreover, $\boldsymbol{n}_{\mathtt{pm}}\in \mathbb{C} ^{ N_{\mathtt{pm}} \times 1}$ denotes an AWGN vector whose entries are i.i.d. $\mathcal{CN}(0,1)$. 
The received signal at $m''$-th S-AP, $m'' \in\mathcal{M}_{s,r}$ can be expressed by:
\vspace{-1mm}
\begin{align}\label{eq:y_m''_xmk}
    &\!\mathbf{y}_{m''} \!=\!\sum\nolimits_{m'\!\in\mathcal{M}_{s,t} } \! (\sqrt{\alpha}\mathbf{h}_{t,m''}\mathbf{h}_{m',t}^{T}\!+\!\mathbf{G}_{m',m''}^{T}\!)\mathbf{x}_{m',t}\notag\\
    &+\sum\nolimits_{m\in\mathcal{M}_c } \mathbf{G}_{m,m''}^{T}\mathbf{x} _{m}  \notag\\
    &+\!(\mathbf{G}_{\mathtt{pm},m''}^{T}+\sqrt{\alpha}\mathbf{h}_{t,m''}\mathbf{h}_{\mathtt{pm},t}^{T})\mathbf{x}_{\mathtt{pm}}+\boldsymbol{n}_{m''},
\end{align}
where $\mathbf{G}_{m,m''}\in \mathbb{C} ^{ N \times N}$ denotes the channel between the $m$-th C-AP and $m''$-th S-AP, while $\boldsymbol{n}_{m''}\in \mathbb{C} ^{ N \times 1}$ represents an AWGN vector whose entries are i.i.d. $\mathcal{CN}(0,1)$. We note that  because all S-APs cooperate,   the AP-AP interference (the terms includes $\mathbf{G}_{m',m''}$)  can be canceled out in~\eqref{eq:y_m''_xmk}. Thus, we can obtain $\!\mathbf{y}_{m''} \!=\!\sum\nolimits_{m'\!\in\mathcal{M}_{s,t} } \!\sqrt{\alpha}\mathbf{h}_{t,m''}\mathbf{h}_{m',t}^{T}\mathbf{x}_{m',t}
    +\sum\nolimits_{m\in\mathcal{M}_c }\!\mathbf{G}_{m,m''}^{T}\mathbf{x} _{m}  
    +\!(\mathbf{G}_{\mathtt{pm},m''}^{T}+\sqrt{\alpha}\mathbf{h}_{t,m''}\mathbf{h}_{\mathtt{pm},t}^{T})\mathbf{x}_{\mathtt{pm}}+\boldsymbol{n}_{m''}$.

\section{Performance Analysis}
In this section, by using the well-known use-and-then-forget bounding technique \cite{ref15}, we provide closed-form expressions for the  overheard SINR of UE $1$  at the monitor, $\mathrm{SINR}_{\mathtt{pm}}$, the effective  SINR for sensing the target, $\mathrm{SINR}_{\mathtt{cpu}}$,  and the effective SINR  at the  UEs, $\mathrm{SINR}_{k}$. It is worth mentioning that  in this section we provide detailed proof for $\mathrm{SINR}_{\mathtt{pm}}$, but skip proofs for $\mathrm{SINR}_{\mathtt{cpu}}$ and $\mathrm{SINR}_{k}$ due to space constraint. However, these proofs can follow a similar methodology.

The signal received at the monitor given by \eqref{eq:yj_xmk} will be used to detect the signal intended for UE 1 (i.e. $s_{1}$). To do this, maximum-ratio combing technique is used. With maximum-ratio combing, the combined signal is $z_{\mathtt{pm}} = \mathbf{w}_{\mathtt{comb},\mathtt{pm}}^{T}\mathbf{y}_{\mathtt{pm}} 
$, where $\mathbf{w}_{\mathtt{comb},\mathtt{pm}} = \left(\sum\nolimits_{m \in\mathcal{M}_c }\sqrt{\eta _{m,1}\rho _{\mathtt{c}}}\mathbf{G}_{m,\mathtt{pm}}^{T}\mathbf {w} _{m,1}^{\mathtt{c}}\right) ^{\ast }.$

\begin{proposition}\label{Theorem2} The received SINR for UE $1$ at
the monitor is given by \eqref{eq:SINR_J}, shown on the top of the next page, where
\begin{figure*}
\begin{equation}
    \mathrm{SINR}_{\mathtt{pm}}=\frac{|\mathbb{E}\left \{\mathrm {DS}_{\mathtt{pm}}\right \}|^{2}}{\mathbb{E}\left \{ |\mathrm{BU}_{\mathtt{pm}}|^{2} \right \}+\mathbb{E}\left \{ |\mathrm{IC}_{{k}',\mathtt{pm}}|^{2} \right \}+\mathbb{E}\left \{ |\mathrm{IS}_{\mathtt{pm}}|^{2} \right \}+\mathbb{E}\left \{ |\mathrm{SI}_{\mathtt{s}}|^{2} \right \}+\mathbb{E}\left \{ |\mathrm{SI}_{\mathtt{c}}|^{2} \right \}+\mathbb{E}\left \{ |\mathrm{n}_{\mathtt{pm}}|^{2} \right \}}, 
    \label{eq:SINR_J}
\end{equation}
\end{figure*}    
\begin{align}
    &\mathbb{E}\left \{\mathrm {DS}_{\mathtt{pm}}\right \}=\sum\nolimits_{m \in\mathcal{M}_c } \eta _{m,1}\rho _{\mathtt{c}}N_{\mathtt{pm}}\beta_{m,\mathtt{pm}}N\gamma_{m,1}, \notag \\
    &\mathbb{E}\left \{ |\mathrm{BU}_{\mathtt{pm}}|^{2} \right \}\!=\!\sum\nolimits_{m \in\mathcal{M}_c }\!\eta _{m,1}\rho _{\mathtt{c}}^2\beta_{m,\mathtt{pm}}\gamma_{m,1}N^2N_{\mathtt{pm}}(1\!+\!N_{\mathtt{pm}})\notag\\
    &\times\left [\!\eta _{m,1}\beta_{m,\mathtt{pm}}\gamma_{m,1}\!+\!\sum\nolimits_{\tilde{m} \in\mathcal{M}_c}\!\eta _{\tilde{m} ,1}\gamma_{\tilde{m} ,1}\beta_{\tilde{m} ,\mathtt{pm}} \right ] \notag\\
    &-\left | \sum\nolimits_{m \in\mathcal{M}_c } \!\eta _{m,1}\rho _{\mathtt{c}}N_{\mathtt{pm}}\beta_{m,\mathtt{pm}}N\gamma_{m,1}\right |^2\!,\notag\\
    &\!\mathbb{E}\!\left \{\! |\mathrm{IC}_{{k}',\mathtt{pm}}\!|^{2} \!\right \}\!\!=\!\!\sum\nolimits_{\!m \in\mathcal{M}_c }\!\sum\nolimits_{{k}'\ne 1}^{K}\!\!\eta _{m,{k'}}\rho _{\mathtt{c}}^2N_{\mathtt{pm}}\!N\gamma_{m,k'}\beta_{m,\mathtt{pm}} \notag\\
    &\!\times\!\!\left [ \! \eta _{m,1}(\!N_{\mathtt{pm}}\!+\!N\!)\!\beta_{m,\mathtt{pm}}\gamma_{m,1}\!+\!\sum\nolimits_{\tilde{m} \in\mathcal{M}_c }\!\eta _{\tilde{m},1}\!N\!\gamma_{\tilde{m},1}\beta_{\tilde{m},\mathtt{pm}} \!\right ]\!, \notag\\
    &\mathbb{E}\left \{ |\mathrm{IS}_{\mathtt{pm}}|^{2} \right \}\!=\!\sum\nolimits_{m\in\mathcal{M}_c }\sum\nolimits_{m'\in\mathcal{M}_{s,t} } \sqrt{\eta _{m'\!,t}} \eta _{m,1}\rho _{\mathtt{s}}\rho _{\mathtt{c}} \notag\\
    &\times\beta_{m,\mathtt{pm}}\gamma _{m,1}\zeta_{m',t}N_\mathtt{pm}N^2\bigg ( \sqrt{\eta _{m'\!,t}} \beta_{m',\mathtt{pm}}\notag\\
    &\!+\!\sqrt{\eta _{m'\!,t}} N\zeta_{\mathtt{pm},t}\zeta_{m',t}\alpha\!+\!\sum\nolimits_{\tilde{m}'\in\mathcal{M}_{s,t} } \!\sqrt{\eta _{\tilde{m}'\!,t}}N\zeta_{\mathtt{pm},t}\zeta_{\tilde{m}',t}\alpha \bigg ),\notag\\
    &\mathbb{E}\left \{ |\mathrm{SI}_{\mathtt{s}}|^{2} \right \}= \sum\nolimits_{m\in\mathcal{M}_{s} }\eta _{\mathtt{pm},t}\rho _{\mathtt{pm}}\eta _{m,1}\rho _{\mathtt{c}}\zeta_{\mathtt{pm},t}\beta_{m,\mathtt{pm}}N_{\mathtt{pm}}^2N\gamma_{m,1}\notag\\
    &\times\big ( \beta_{\mathtt{pm},\mathtt{pm}}+\alpha N_{\mathtt{pm}}\zeta_{\mathtt{pm},t}^2 \big ),\notag\\
    &\mathbb{E}\left \{\! |\mathrm{SI}_{\mathtt{c}}|^{2} \!\right \}\!=\!\sum\nolimits_{m \in\mathcal{M}_c }\eta _{\mathtt{pm},1}\eta _{m,1}\rho _{\mathtt{c}}\rho _{\mathtt{pm}}\gamma_{m,1}NN_{\mathtt{pm}}^2\beta_{m,\mathtt{pm}}\beta_{\mathtt{pm},1}\notag\\&\times(\beta_{\mathtt{pm},\mathtt{pm}}+\alpha\zeta_{\mathtt{pm},t}^2), \notag\\
    &\mathbb{E}\left \{ |\mathrm{n}_{\mathtt{pm}}|^{2} \right \}=\sum\nolimits_{m\in\mathcal{M}_c }\eta _{m,1}\rho _{\mathtt{c}}NN_{\mathtt{pm}}\beta_{m,\mathtt{pm}}\gamma_{m,1}.
\end{align}
\end{proposition}

\vspace{-1.2mm}
\begin{proof}
See Appendix~\ref{ProofTheorem2}.
\end{proof}

 \begin{proposition}\label{Theorem1}
The effective SINR of the $k$-th UE is given by \eqref{eq:SINR_UE}, shown on the top of the next page, where
\begin{figure*}
\vspace{-0.4cm}
\begin{equation}
\mathrm{SINR}_{k}=\frac{\left | \mathbb{E}\left \{ \mathrm {DS}_{k}\right \}  \right |^2 }{\mathbb{E}\left \{ |\mathrm{BU}_{k}|^{2} \right \}+ \mathbb{E}\left \{|\mathrm{IUI}_{k',k}|^{2}\right \}+\mathbb{E}\left \{ |\mathrm{IS}_{k}|^{2} \right \}+\mathbb{E}\left \{ |\mathrm {JS}_{\mathtt{s},k}|^2 \right\}+\mathbb{E}\left \{ |\mathrm {JS}_{\mathtt{c},k}|^2 \right\}+1},
\label{eq:SINR_UE}
\end{equation}
\hrulefill
\vspace{-0.4cm}
\end{figure*}
\vspace{-1mm}
\begin{align}
    &\mathbb{E}\left \{ \mathrm {DS}_{k}\right \}  =  \sum\nolimits_{m \in\mathcal{M}_c }\sqrt{\eta _{m,k}\rho _{\mathtt{c}}}N\gamma _{m,k},\notag\\
    &\mathbb{E}\left \{ |\mathrm{BU}_{k}|^{2} \right \} 
    =\rho _{\mathtt{c}}N \sum\nolimits_{m \in\mathcal{M}_c }\eta _{m,k}\gamma _{m,k}\beta_{m,k},\notag\\
    &\mathbb{E}\left \{ |\mathrm{IUI}_{k',k}|^{2} \right \}\! =  \!\sum\nolimits_{m \in\mathcal{M}_c }\sum\nolimits_{{k}'\ne k}^{K}\eta _{m,k'}\rho _{\mathtt{c}}N\gamma_{m,k'}\beta_{m,k}, \notag\\
    &\mathbb{E}\left \{ |\mathrm{IS}_{k}|^{2} \right \} \!=\!\sum\nolimits_{m'\in\mathcal{M}_{s,t} }\sqrt{\eta _{m'\!,t}}\rho _{\mathtt{s}}\zeta_{m',t}N\notag\\
    &\!\times\!\bigg(\beta_{m',k}\!+\!\alpha\zeta_{m',t}\zeta_{t,k}N\!+\!\sum\nolimits_{\tilde{m}'\in\mathcal{M}_{s,t} }\!\sqrt{\eta _{\tilde{m}'\!,t}}\zeta_{\tilde{m}',t}\zeta_{t,k}\alpha N\bigg), \notag\\
    &\mathbb{E}\left \{ |\mathrm {JS}_{\mathtt{s},k}|^2 \right\} =\eta _{\mathtt{pm},t}\rho _{\mathtt{pm}} (\beta_{\mathtt{pm},k}\zeta_{\mathtt{pm},t}N_{\mathtt{pm}}+\alpha\zeta_{t,k}N_{\mathtt{pm},t}^2\zeta_{\mathtt{pm},t}^2), \notag\\
    &\mathbb{E}\left \{ |\mathrm {JS}_{\mathtt{c},k}|^2 \right\}\!=\!
    \eta_{\mathtt{pm},\!1}\rho _{\mathtt{pm}}N_{\mathtt{pm}}\beta_{\mathtt{pm},1}(N_{\mathtt{pm}}\beta_{\mathtt{pm},1}\!+\!\beta_{\mathtt{pm}\!,1}\!+\!\alpha\zeta_{t\!,k}\zeta_{\mathtt{pm},t}).
\end{align}
\end{proposition}
\vspace{-3mm}
 
Using the combining vector $\mathbf{w}_{\mathtt{comb},m''} = \big(\sum\nolimits_{m'\in\mathcal{M}_{s,t} } \sqrt{\eta _{m',t}\rho _{\mathtt{s}}}\sqrt{\alpha}\mathbf{h}_{t,m''}\mathbf{h}_{m',t}^{T}\mathbf{w}_{m',t}^{\mathtt{s}}\big) ^{\ast}$, 
at the $m''$-th S-AP, 
the received  signal at the CPU of the malicious ISAC system to detect the target can be expressed by ${z}_{\mathtt{cpu}}=\sum\nolimits_{m''\in\mathcal{M}_{s,r} }\!\mathbf{w}_{\mathtt{comb},m''}^{T}\mathbf{y}_{m''}$.
Accordingly, the received SINR at the CPU for sensing
the target  can be obtained as in the following
proposition.

\begin{proposition}~\label{Theorem3}
The received $\mathrm{SINR}$ at the CPU for sensing the target is given by \eqref{eq:SINR_m''} on the top of the next page, where
\begin{figure*}
\begin{equation}
 \mathrm{SINR}_{\mathtt{cpu}}=\frac{|\mathbb{E}\left \{\mathrm {DS}_{\mathtt{cpu}}\!\right \}|^{2}}{\mathbb{E}\left \{ |\mathrm{BU}_{\mathtt{cpu}}|^{2} \right \}+\mathbb{E}\left \{ |\mathrm{IC}_{\mathtt{cpu}}|^{2} \right \} +\mathbb{E}\left \{ |\mathrm{JS}_{\mathtt{s},\mathtt{cpu}}|^{2} \right \}+\mathbb{E}\left \{ |\mathrm{JS}_{\mathtt{c},\mathtt{cpu}}|^{2} \right \}+\mathbb{E}\left \{ |\mathrm{n}_{\mathtt{cpu}}|^{2} \right \}},
\label{eq:SINR_m''}
\end{equation}
\hrulefill
\vspace{-0.3cm}
\end{figure*}
\begin{align}
    &\!\mathbb{E}\left \{\mathrm {DS}_{\mathtt{cpu}}\!\right \}= \!\sum\nolimits_{m''\!\in\mathcal{M}_{s,r} }\sum\nolimits_{m'\in\mathcal{M}_{s,t} } \sqrt{\eta _{m'\!,t}}\rho_{\mathtt{s}} \zeta_{m',t}\zeta_{t,m''}\notag\\
    &\times\alpha N^3\bigg ( \sqrt{\eta _{m'\!,t}} \zeta_{m',t}+\sum\nolimits_{\tilde{m}'\in\mathcal{M}_{s,t} } \sqrt{\eta _{\tilde{m}'\!,t}}\zeta_{\tilde{m}',t} \bigg ), \notag\\
    &\mathbb{E}\left \{\! |\mathrm{BU}_{\mathtt{cpu}}|^{2} \! \right \}\!=0,\notag \\
    &\mathbb{E}\left \{\! |\mathrm{IC}_{\mathtt{cpu}}|^{2}\! \right \}\!= \!\sum\nolimits_{m''\!\in\mathcal{M}_{s,r} }\!\sum\nolimits_{m\in\mathcal{M}_c }\! \sum\nolimits_{k =1}^{K}\!\sum\nolimits_{m'\!\in\mathcal{M}_{s,t} }\!\!\eta _{m,k} \notag\\
    &\times\sqrt{\eta _{m',t}}\rho _{\mathtt{c}}\rho _{\mathtt{s}}\gamma_{m,k}N^4\! \beta_{m,m''} \zeta_{t,m''}\zeta_{m',t}\notag\\
    &\times\bigg( \sqrt{\eta _{m',t}}\zeta_{m',t}+ \sum\nolimits_{\tilde{m}'\!\in\mathcal{M}_{s,t} }\sqrt{\eta _{\tilde{m}',t}}\zeta_{\tilde{m}',t}\bigg ) , \notag\\
    &\mathbb{E}\!\left \{\! |\mathrm{JS}_{\mathtt{s},\mathtt{cpu}}|^{2} \right \}=\!\sum\nolimits_{m''\!\in\mathcal{M}_{s,r} }\!\sum\nolimits _{m'\!\in\mathcal{M}_{s,t} }\sqrt{\eta_{m',t}}\eta _{\mathtt{pm},t}\rho _{\mathtt{s}}\rho _{\mathtt{pm}}\alpha \notag\\
    &\times N^3N_{\mathtt{pm}}\zeta_{\mathtt{pm},t}\zeta_{t,m''}\zeta_{m',t}(\beta_{\mathtt{pm},m''}\!+\!\alpha \zeta_{\mathtt{pm},t}\zeta_{t,m''}NN_{\mathtt{pm}})\notag\\
    &\times\bigg( \sqrt{\eta _{m',t}}\zeta_{m',t}+ \sum\nolimits_{\tilde{m}'\!\in\mathcal{M}_{s,t} }\sqrt{\eta _{\tilde{m}',t}}\zeta_{\tilde{m}',t}\bigg ), \notag\\
    &\mathbb{E}\!\left \{\! |\mathrm{JS}_{\mathtt{c},\mathtt{cpu}}|^{2}\! \right \}\!=\sum\nolimits_{m''\!\in\mathcal{M}_{s,r} }\sum\nolimits_{m'\!\in\mathcal{M}_{s,t} }\sqrt{\eta _{m',t}}\eta _{\mathtt{pm},1}\rho _{\mathtt{pm}}\rho _{\mathtt{s}} \notag\\
    &\times \alpha \zeta_{t,m''}\zeta_{m',t}N_{\mathtt{pm}}N^3\beta_{\mathtt{pm},1}(\beta_{\mathtt{pm},m''}+\alpha N\zeta_{\mathtt{pm},t}\zeta_{t,m''})\notag\\
    &\times \bigg(\sqrt{\eta _{m'\!,t}} \zeta_{m',t}+\sum\nolimits_{\tilde{m}'\in\mathcal{M}_{s,t} }\sqrt{\eta _{\tilde{m}'\!,t}} \zeta_{\tilde{m}',t}\bigg),\notag\\
    &\mathbb{E}\left \{ |\mathrm{n}_{\mathtt{cpu}}|^{2} \right \}\!=\! \sum\nolimits_{m''\!\in\mathcal{M}_{s,r} }\!\sum\nolimits_{m'\!\in\mathcal{M}_{s,t} }\!\sqrt{\eta _{m',t}}\rho _{\mathtt{s}}\alpha \zeta_{t,m''}\zeta_{m',t} \notag\\
    &\times N^3\bigg( \sqrt{\eta _{m',t}}\zeta_{m',t}+ \sum\nolimits_{\tilde{m}'\!\in\mathcal{M}_{s,t} }\sqrt{\eta _{\tilde{m}',t}}\zeta_{\tilde{m}',t}\bigg ).
\end{align}
\end{proposition}
\vspace{-1mm}
A suitable performance metric for monitoring the malicious UE  is the  MSP.  If $\mathrm{SINR}_{\mathtt{pm}}\ge \mathrm{SINR}_1$, the monitor can  reliably detect the information of  UE $1$. On the other hand, if $\mathrm{SINR}_{\mathtt{pm}}<\mathrm{SINR}_1$, the monitor may detect this information with error. Therefore, the following indicator function can be considered for characterizing the event of successful monitoring at the monitor:
\vspace{-2mm}
\begin{equation}
    X_1= \left\{\begin{matrix}
 1, & \mathrm{SINR}_{\mathtt{pm}}\ge \mathrm{SINR}_1\\
 0, & \mathrm{SINR}_{\mathtt{pm}}<  \mathrm{SINR}_1.
\end{matrix}\right.
\vspace{-1mm}
\label{eq:X_k}
\end{equation}
Therefore,  the MSP $\mathbb{E}\left \{ X_1 \right \}$ can be written as $\mathbb{E}\left \{ X_1 \right \} = \mathrm {Pr} \left \{ \mathrm{SINR}_{\mathtt{pm}}\ge \mathrm{SINR}_1 \right \} $ \cite{ref:zahra_iot}. 
In addition, to evaluate the effect of monitor on the sensing performance of  the malicious ISAC, we consider the  SDP of the target by the S-APs.  We highlight that there is a cooperation between the S-APs which can be handled by the CPU. The SDP of the legitimate target
can be written as $\mathrm {Pr}  \left \{ \mathrm{SINR}_{\mathtt{cpu}}  \ge \kappa \right \}${~\cite{ref:jj_He_localization},} where $ \kappa $ is the minimum SINR required for successful detection.
\begin{figure}[t]
  \centering
  \includegraphics[width=0.85\linewidth]{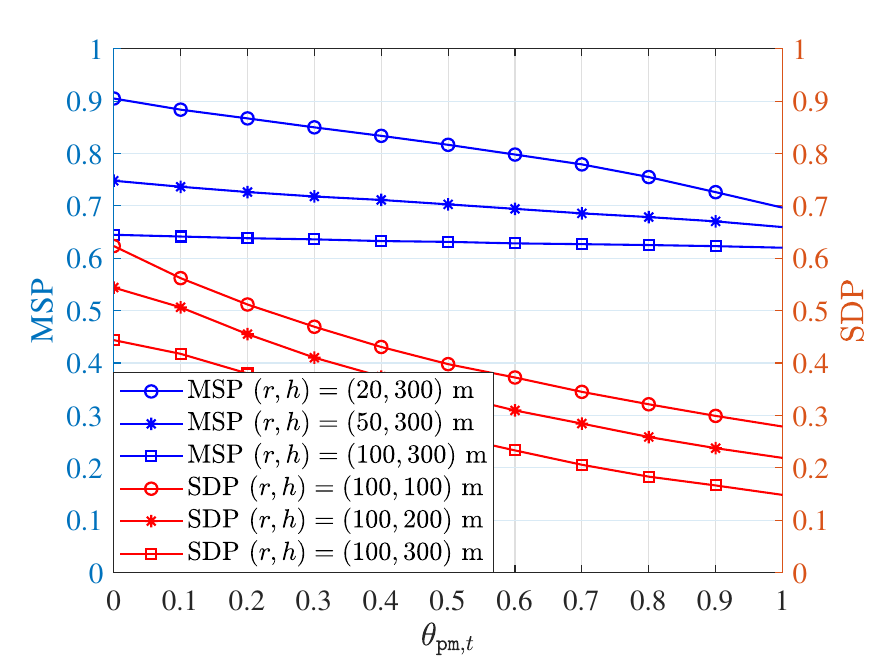}
  \caption{MSP and SDP versus $\theta_{\mathtt{pm},t}$. }
  \label{fig:monitoring_success_probability}
\end{figure}
\section{Simulation Results}
In this section, numerical results are presented to investigate the performance of our anti-malicious CF-mMIMO ISAC design and to evaluate the accuracy of the derived SINR expressions. We assume that each AP is equipped with a uniform linear array (ULA) antenna. The $n$-th element of the steering vector $\boldsymbol{\alpha }_{t}\left( \phi_{m',t}^{a}, \phi_{m',t}^{e}\right) \in \mathbb{C}^{N\times 1}$ is
$\left [\boldsymbol{\alpha } _{t}\left ( \phi_{m',t}^{a}, \phi_{m',t}^{e}\right ) \right ]_{n}\!=\!{\exp\left [ j2\pi\frac{\Delta d}{\lambda }(n-1) \sin \phi_{m',t}^{a}\cos \phi_{m',t}^{e} \right ] }$, where $\Delta d= \frac{\lambda}{2}$ is the distance between any two adjacent antennas.
We consider $20$ C-APs for transmitting communication signals, $3$  APs for transmitting and $3$  APs for receiving sensing signals, and $5$  UEs in the malicious ISAC system. The APs and UEs are randomly distributed in an area of $1\times 1$ $\mathrm{km}^2$ having wrapped around edges to reduce the boundary effects. The legitimate monitor is positioned randomly on a circle centred around UE $1$ with a radius of $r$, while the target is randomly located in an area at the altitude of $h$ above the ground. 
We select a three-slope path-loss model and calculate the path loss 
$\mathrm {PL}_{m,k}$ following the modeling approach of~\cite{ref14}. The sensing SINR threshold $\kappa$ is $3$ dB and the RCS of target is $\sigma _{RCS} = 0.1$ $\mathrm{m}^2$ \cite{ref:CYu_cellfree_ISAC}. 

The large-scale fading coefficient can be calculated by $\beta_{m,k} = \mathrm {PL}_{mk} \cdot 10^{\frac{\sigma _{sh}z_{mk}}{10} }  $.
Hereby, we choose $\sigma_{sh}=9$dB and $z_{mk}\sim \mathcal{CN} (0,1)$. We further consider that the noise figure $\sigma_{n}$ equals to $8$ dB. Moreover, we assume that the power for downlink data transmission and sensing $P _{c} = P_{\mathtt{s}} = 1$ W, the transmit power for sending the pilot sequences is $P _{p} =0.2$ W and the power allocated at the monitor is denoted by $P_{\mathtt{pm}}$,
while the  normalized maximum transmit powers $\rho _{\mathtt{c}}$, $\rho _{\mathtt{s}}$, $\rho _{p}$ and $\rho _{\mathtt{pm}}$ can be calculated upon dividing the corresponding powers by the noise power. Considering that the APs transmit at full power, the transmit power coefficient at C-APs and S-APs can be expressed by $ \eta_{m,k} = \frac{1}{{N\sum_{k=1}^{K} \gamma_{m,k}}}$ and $\eta _{m'\!,t} = \frac{1}{N\zeta_{m',t}}$, respectively~\cite{ref9}.

\begin{figure}[t]
  \centering
  \includegraphics[width=0.85\linewidth]{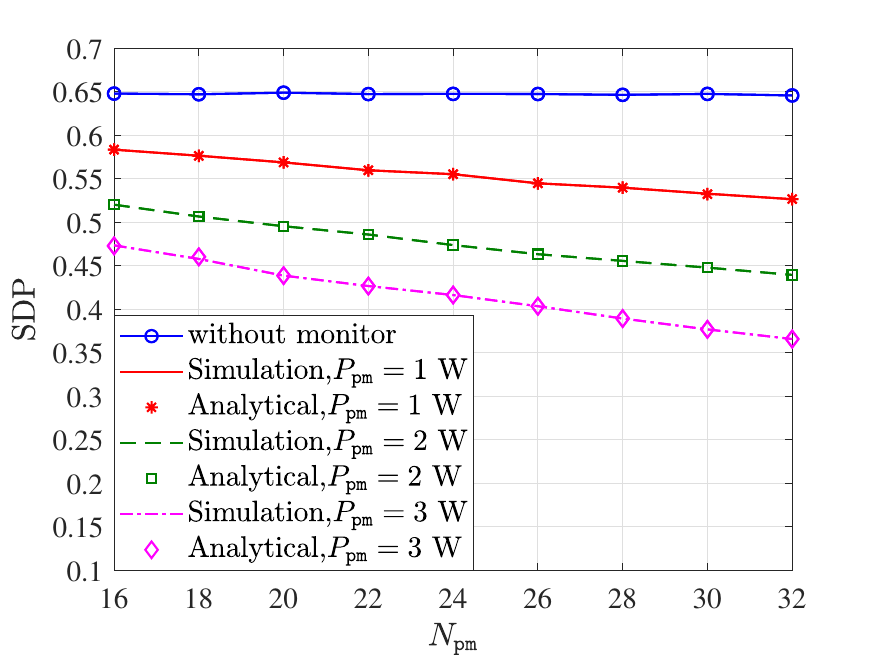}
  \caption{SDP versus $N_\mathrm{pm}$.}
  \label{fig:success_detection_rate}
\end{figure}
In addition, the power constraint at the monitor can be expressed as 
$\theta_{\mathtt{pm},t}+\theta_{\mathtt{pm},1}\le 1$,
where  $\theta_{\mathtt{pm},t} \triangleq N_{\mathtt{pm}}\eta _{\mathtt{pm},t}\zeta_{\mathtt{pm},t}$   and  $\theta_{\mathtt{pm},1}\triangleq N_{\mathtt{pm}}\eta _{\mathtt{pm},1}\beta_{\mathtt{pm},1}$. We note that $\rho_\mathtt{pm}\theta_{\mathtt{pm},t}$   represents the power allocated at the monitor to jam the target, while $\rho_\mathtt{pm}\theta_{\mathtt{pm},1}$   represents the power allocated at the monitor to jam UE $1$.

Figure 2 shows the MSP and SDP  as a function of $\theta_{\mathtt{pm},t}$, where $N=5$ and $N_{\mathrm{pm}}=32$. 
We first observe that the SDP of the malicious ISAC system  reduces significantly with $\theta_{\mathtt{pm},t}$. 
The reason is that as $\theta_{\mathtt{pm},t}$ increases, the reflected jamming signal due to the monitor's jamming transmission to the target causes stronger interference on the S-APs, which degrades
the received SINR at the S-APs, and hence deteriorates the SDP performance. In addition, we can see that the MSP notably reduces  with  $\theta_{\mathtt{pm},t}$  when $r$ is low. This is reasonable, because by increasing $\theta_{\mathtt{pm},t}$, on one hand, the reflected jamming interference at UE $1$, due to the monitor's jamming transmission to the target, becomes stronger which degrades
the received SINR at UE $1$. On the other hand, as $\theta_{\mathtt{pm},t}$ increases $\theta_{\mathtt{pm},1}$ decreases, and hence the jamming interference at UE $1$  decreases. The latter effect is dominant for small to medium values of $r$, which reduces the MSP performance.  Figure 2 confirms that the proposed anti-malicious ISAC system can successfully monitor the malicious ISAC system, providing excellent MSP performance, while substantially decreasing the  SDP by around $66\%$.  Therefore, by carefully selecting the power allocation coefficients $\theta_{\mathtt{pm},t}$ and  $\theta_{\mathtt{pm},1}$ at the monitor, our proposed anti-malicious ISAC design can provide an acceptable trade-off between the SDP and MSP.

Figure 3 illustrates the  SDP of the malicious CF-mMIMO ISAC system versus $N_\mathrm{pm}$ with and without monitor. Also, $\eta _{\mathtt{pm},t}=\frac{1}{2 N_{\mathtt{pm}}\zeta_{\mathtt{pm},t}}$ and $\eta _{\mathtt{pm},1}=\frac{1}{2 N_{\mathtt{pm}}\beta_{\mathtt{pm},1}}$. Our simulation results lead to the following
conclusions: i) the introduction of a monitor in our anti-malicious ISAC scheme can significantly hinder the sensing capability of malicious ISAC system, e.g., $46\%$  decrease in the SDP performance   when $P_{\mathtt{pm}}=3$ W and $N_\mathtt{pm}=32$; ii) the SDP decreases significantly with $N_{\mathtt{pm}}$, especially when $P_{\mathrm{pm}}$  is high; iii) the analytical results (solid curves) match tightly with the simulation results (markers).
 \vspace{-0.1cm}
\section{Conclusion}
This paper proposed an anti-malicious CF-mMIMO ISAC  via
proactive monitoring approach to efficiently intercept the suspicious communication link and compromise the sensing performance. We deduced closed-form expressions for the SINR at the S-APs, UEs, and monitor. The numerical results confirmed the accuracy of our analysis and showed that the introduction of a proactive monitor can bring a decrease of $66\%$ in the SDP performance of the malicious ISAC, while providing excellent MSP performance. Future work will include proactive monitoring relying on power optimization, extending the results to a multi-target setup, and also look into how to acquire channel state information (CSI) of the malicious ISAC system in practice.



%

\vspace{-01cm}
\appendices
\section{Proof of Proposition~\ref{Theorem2}}\label{ProofTheorem2}
\vspace{-0.1cm}

1) Compute $\mathrm {DS}_{\mathtt{pm}}$:
\vspace{-0.1cm}
\begin{align}\label{eq:DS_J}
   &\mathbb{E}\left \{\mathrm {DS}_{\mathtt{pm}}\right \}
    =\sum\nolimits_{m \in\mathcal{M}_c } \sqrt{\eta _{m,1}\rho _{\mathtt{c}}} \mathbb{E}\left \{\mathbf{w}_{\mathtt{comb},\mathtt{pm}}^{T}\mathbf{G}_{m,\mathtt{pm}}^{T}\mathbf {w} _{m,1}^{\mathtt{c}} \right \} \notag\\
    &=\sum\nolimits_{m \in\mathcal{M}_c } \eta _{m,1}\rho _{\mathtt{c}}\mathbb{E}\left \{\hat{\mathbf{g}}_{m,1}^{T}\mathbf{G}_{m,\mathtt{pm}}^{\ast}\mathbf{G}_{m,\mathtt{pm}}^{T}\hat{\mathbf{g}}_{m,1}^{\ast} \right \},\notag\\
    &=\sum\nolimits_{m \in\mathcal{M}_c } \eta _{m,1}\rho _{\mathtt{c}}N_{\mathtt{pm}}\beta_{m,\mathtt{pm}}N\gamma_{m,1}.
\end{align}

2)  Using \cite[Lemma 10]{ref4}, the beamforming uncertainty $\mathbb{E}\left \{ |\mathrm{BU}_{\mathtt{pm}}|^{2} \right \}=\mathbb{E}\bigg \{ \!\bigg |\mathbf{w}_{\mathtt{comb},\mathtt{pm}}^{T}\!\sum\nolimits_{m \in\mathcal{M}_c } \!\sqrt{\eta _{m,1}\rho _{\mathtt{c}}}\mathbf{G}_{m,\mathtt{pm}}^{T}\mathbf {w} _{m,1}^{\mathtt{c}}- \mathbb{E}\big \{\mathrm {DS}_{\mathtt{pm}}\big \}\bigg|^2\bigg \}$ can be computed as (\ref{eq:BU_J}).
\begin{figure*}
\begin{align}\label{eq:BU_J}
    &\mathbb{E}\left \{ |\mathrm{BU}_{\mathtt{pm}}|^{2} \right \}\!
    =\! \mathbb{E}\Big \{ \!\Big |\mathbf{w}_{\mathtt{comb},\mathtt{pm}}^{T}\!\sum\nolimits_{m \in\mathcal{M}_c } \!\sqrt{\eta _{m,1}\rho _{\mathtt{c}}}\mathbf{G}_{m,\mathtt{pm}}^{T}\mathbf {w} _{m,1}^{\mathtt{c}}\Big|^2\Big \}\!-\!\Big |  \mathbb{E}\Big \{ \!\mathbf{w}_{\mathtt{comb},\mathtt{pm}}^{T}\!\sum\nolimits_{m \in\mathcal{M}_c } \!\sqrt{\eta _{m,1}\rho _{\mathtt{c}}}\mathbf{G}_{m,\mathtt{pm}}^{T}\mathbf {w} _{m,1}^{\mathtt{c}}\Big \}\Big |^2, \notag\\
    &= \mathbb{E}\left \{ |(\sum\nolimits_{m \in\mathcal{M}_c }\sqrt{\eta _{m,1}\rho _{\mathtt{c}}}\mathbf{G}_{m,\mathtt{pm}}^{T}\hat{\mathbf{g} }_{m,1} ^{\ast }) ^{H }\sqrt{\eta _{m,1}\rho _{\mathtt{c}}}\mathbf{G}_{m,\mathtt{pm}}^{T}\hat{\mathbf{g} }_{m,1} ^{\ast }|^2\right \}-\left | \sum\nolimits_{m \in\mathcal{M}_c } \eta _{m,1}\rho _{\mathtt{c}}N_{\mathtt{pm}}\beta_{m,\mathtt{pm}}N\gamma_{m,1}\right |^2 \notag\\
    &+\mathbb{E}\left \{ |(\sum\nolimits_{m \in\mathcal{M}_c }\sqrt{\eta _{m,1}\rho _{\mathtt{c}}}\mathbf{G}_{m,\mathtt{pm}}^{T}\hat{\mathbf{g} }_{m,1} ^{\ast }) ^{H } \sum\nolimits_{\tilde{m} \in\mathcal{M}_c } \!\sqrt{\eta _{\tilde{m},1}\rho _{\mathtt{c}}}\mathbf{G}_{\tilde{m} ,\mathtt{pm}}^{T}\hat{\mathbf{g} }_{\tilde{m} ,1} ^{\ast }|^2\right \}-0, \notag \\
    &\!=\!\sum_{m \in\mathcal{M}_c }\!\eta _{m,1}\rho _{\mathtt{c}}^2\beta_{m,\mathtt{pm}}\gamma_{m,1}N^2N_{\mathtt{pm}}\!(N_{\mathtt{pm}}\!+\!1)\!\Big [ \eta _{m,1}\!\beta_{m,\mathtt{pm}}\gamma_{m,1}\!+\!\sum_{\tilde{m} \in\mathcal{M}_c }\!\eta _{\tilde{m} ,1}\gamma_{\tilde{m} ,1}\beta_{\tilde{m} ,\mathtt{pm}} \Big ] \!-\!\Big |\! \sum_{m \in\mathcal{M}_c } \!\eta _{m,1}\rho _{\mathtt{c}}N_{\mathtt{pm}}\beta_{m,\mathtt{pm}}N\gamma_{m,1}\Big |^2\!.
\end{align}
\hrulefill
\vspace{-4mm}
\end{figure*}

3) Compute $\mathbb{E}\left \{ |\mathrm{IC}_{{k}',\mathtt{pm}}|^{2} \right \}$:
    \vspace{-3mm}
\begin{align}\label{eq:IC_k',J}
    &\!\!\mathbb{E}\big \{\! |\mathrm{IC}_{{k}',\mathtt{pm}}\!|^{2} \!\big \}
    \!\!=\!\!\mathbb{E}\bigg \{\!\bigg | \mathbf{w}_{\!\mathtt{comb},\mathtt{pm}}^T\!\!\!\sum_{m \in\mathcal{M}_c }\! \sum_{{k}'\ne 1}^{K}\!\sqrt{\eta _{m,{k'}}\!\rho _{\mathtt{c}}}\mathbf{G}_{m,\mathtt{pm}}^{T}\hat{\mathbf{g} }_{m,k'} ^{\ast }\! \bigg |^{2}\! \bigg \}, \notag\\
    &\!\!=\!\!\!\sum_{m \in\mathcal{M}_c }\!\sum_{{k}'\ne 1}^{K}\!\!\eta _{m,1}\eta _{m,{k'}}\rho _{c}^2\gamma_{m,k'}\mathbb{E}\!\big \{ \hat{\mathbf{g} }_{m,\!1}^{T}\mathbb{E}\!\big \{\! \big | \mathbf{G}_{m,\mathtt{pm}}^{\ast } \!\mathbf{G}_{m,\mathtt{pm}}^{T}  \!\big |^2\!\big \}\hat{\mathbf{g} }_{m,\!1}^{\ast } \!\big \}\notag\\
    &\!\!+\!\!\sum_{m \in\mathcal{M}_c }\!\sum_{{k}'\ne 1}^{K}\!\sum_{\tilde{m} \in\mathcal{M}_c}\!\!\eta _{m,{k'}}\eta _{\tilde{m},1}\rho _{c}^2N\!\gamma_{m,k'}\beta_{m,\mathtt{pm}}\mathbb{E}\!\big \{\big | \hat{\mathbf{g} }_{\tilde{m},\!1}^{T}\!\mathbf{G}_{\tilde{m},\mathtt{pm}}^{*} \!\big |^{2} \!\big \}, \notag\\
    &\!\!=\!\!\sum_{m \in\mathcal{M}_c }\!\sum_{{k}'\ne 1}^{K}\!\!\eta _{m,\!1}\eta _{m,{k'}}\rho _{c}^2\!\gamma_{m,k'}\mathbb{E}\!\big \{ \hat{\mathbf{g} }_{m\!,1}^{T}\beta_{m,\mathtt{pm}}^2N_{\mathtt{pm}}\!(N\!\!+\!\!N_{\mathtt{pm}})\hat{\mathbf{g} }_{m,1}^{\ast } \!\big \}\notag\\
    &\!\!+\!\!\sum_{m \in\mathcal{M}_c }\!\sum_{{k}'\ne 1}^{K}\!\sum_{\tilde{m} \in\mathcal{M}_c}\!\!\eta _{m,{k'}}\eta _{\tilde{m},1}\rho _{c}^2N^2N_{\mathtt{pm}}\gamma_{m,k'}\gamma_{\tilde{m},1}\beta_{m,\mathtt{pm}}\beta_{\tilde{m},\mathtt{pm}},\notag\\
    &\!=\!\sum_{m \in\mathcal{M}_c }\!\!\sum_{{k}'\ne 1}^{K}\eta _{m,{k'}}\rho _{\mathtt{c}}^2N_{\mathtt{pm}}\!N\gamma_{m,k'}\beta_{m,\mathtt{pm}}\notag\\
    &\!\times\!\bigg[  \eta _{m,1}(\!N_{\mathtt{pm}}\!+\!N\!)\beta_{m,\mathtt{pm}}\gamma_{m,1}\!+\!\sum_{\tilde{m} \in\mathcal{M}_c }\!\!\eta _{\tilde{m},1}N\gamma_{\tilde{m},1}\beta_{\tilde{m},\mathtt{pm}} \bigg ].
\end{align}

4) Compute $\mathbb{E}\left \{ |\mathrm{IS}_{\mathtt{pm}}|^{2} \right \}$: The derivation is shown in (\ref{eq:IS_j}).
\begin{figure*}
\vspace{-0.1cm}
\begin{align}\label{eq:IS_j}
        &\!\mathbb{E}\!\big \{ |\mathrm{IS}_{\mathtt{pm}}|^{2} \big \}\!=\! \mathbb{E}\!\Big\{ \!\Big |\mathbf{w}_{\mathtt{comb},\mathtt{pm}}^T\!\sum\nolimits_{m'\in\mathcal{M}_{s,t} } \!\sqrt{\eta _{m'\!,t}\rho _{\mathtt{s}}}\mathbf{G}_{m',\mathtt{pm}}^{T}\mathbf{h}_{m',t}^{\mathtt{\ast}}  \Big |^2 \Big \}+\mathbb{E}\Big \{ \Big |\mathbf{w}_{\mathtt{comb},\mathtt{pm}}^T\sum\nolimits_{m'\in\mathcal{M}_{s,t} } \sqrt{\eta _{m'\!,t}\rho _{\mathtt{s}}}\sqrt{\alpha}\mathbf{h}_{\mathtt{pm},t}\mathbf{h}_{m',t}^{T}\mathbf{h}_{m',t}^{\mathtt{\ast}}  \Big |^2 \Big \}, \notag\\
        &=\sum\nolimits_{m\in\mathcal{M}_c }\sum\nolimits_{m'\in\mathcal{M}_{s,t} } \eta _{m'\!,t}\rho _{\mathtt{s}}\eta _{m,1}\rho _{\mathtt{c}}N\beta_{m',\mathtt{pm}}\zeta_{m',t}\mathbb{E}\left \{\hat{\mathbf{g}}_{m,1}^{T}\mathbf{G}_{m,\mathtt{pm}}^{\ast}\mathbf{I}_{\mathtt{pm}}\mathbf{G}_{m,\mathtt{pm}}^{T}\hat{\mathbf{g}}_{m,1}^{\ast}\right \}\notag\\
       &+ \sum\nolimits_{m \in\mathcal{M}_c }\sum\nolimits_{m'\in\mathcal{M}_{s,t} }\eta _{m'\!,t}\rho _{\mathtt{s}} \eta _{m,1}\rho _{\mathtt{c}}N^2N_{\mathtt{pm}}\zeta_{\mathtt{pm},t}\zeta_{m',t}^2\beta_{m,\mathtt{pm}}\alpha\mathbb{E}\left \{ \left | \hat{\mathbf{g} }_{m,1} ^{T}\! \right |^2 \right \}\notag\\
       &+\sum\nolimits_{m \in\mathcal{M}_c }\sum\nolimits_{m'\in\mathcal{M}_{s,t} }\sqrt{\eta _{m'\!,t}}\rho _{\mathtt{s}} \eta _{m,1}\rho _{\mathtt{c}}N^3N_{\mathtt{pm}}\zeta_{\mathtt{pm},t}\zeta_{m',t}\beta_{m,\mathtt{pm}}\alpha\gamma_{m,1}\sum\nolimits_{\tilde{m}'\in\mathcal{M}_{s,t} }\sqrt{\eta _{\tilde{m}'\!,t}}\zeta_{\tilde{m}',t}\notag\\
        &\!=\!\!\sum_{m\in\mathcal{M}_c }\!\sum_{m'\in\mathcal{M}_{s,t} }\! \!\sqrt{\eta _{m'\!,t}} \eta _{m,1}\rho _{\mathtt{s}}\rho _{\mathtt{c}}\beta_{m,\mathtt{pm}}\gamma _{m,1}\zeta_{m',t}N_\mathtt{pm}N^2 \Big ( \sqrt{\eta _{m'\!,t}} \beta_{m',\mathtt{pm}}\!+\!\sqrt{\eta _{m'\!,t}} N\zeta_{\mathtt{pm},t}\zeta_{m',t}\alpha\!+\!\!\sum_{\tilde{m}'\in\mathcal{M}_{s,t} }\! \!\sqrt{\eta _{\tilde{m}'\!,t}}N\zeta_{\mathtt{pm},t}\zeta_{\tilde{m}',t}\alpha \Big ).
\end{align}
\vspace{-0.1cm}
\hrulefill
\vspace{-0.4cm}
\end{figure*}

5) Compute $\mathbb{E} \{ |\mathrm{SI}_{\mathtt{s}}|^{2}  \} $:
\vspace{-1em}
\begin{align}
   &\mathbb{E}\big \{\! |\mathrm{SI}_{\mathtt{s}}|\!^{2} \big \}=\mathbb{E}\big \{ \big |\mathbf{w}_{\mathtt{comb},\mathtt{pm}}^T\sqrt{\eta _{\mathtt{pm},t}\rho _{\mathtt{pm}}}\mathbf{G}_{\mathtt{pm},\mathtt{pm}}^{T}\mathbf {h}_{\mathtt{pm},t}^{\ast}\big |^2 \big \}\notag\\
   &+\mathbb{E}\big \{ \big |\mathbf{w}_{\mathtt{comb},\mathtt{pm}}^T\sqrt{\eta _{\mathtt{pm},t}\rho _{\mathtt{pm}}}\sqrt{\alpha}\mathbf{h}_{t,\mathtt{pm}}\mathbf{h}_{\mathtt{pm},t}^{T}\mathbf {h}_{\mathtt{pm},t}^{\ast}\big |^2 \big \},\notag  \\
   &\!=\!\sum\nolimits_{m \in\mathcal{M}_c }\eta _{m,1}\rho _{\mathtt{c}}\eta _{\mathtt{pm},t}\rho _{\mathtt{pm}}N_{\mathtt{pm}}\beta_{\mathtt{pm},\mathtt{pm}}\zeta_{\mathtt{pm},t}\mathbb{E}\big \{ \big |\hat{\mathbf{g} }_{m,1}^{T}\mathbf{G}_{m,\mathtt{pm}}^{\ast }\big |^2 \big \}\notag\\
   &\!+\!\sum\nolimits_{m \in\mathcal{M}_c }\!\eta _{m,1}\rho _{\mathtt{c}}\eta _{\mathtt{pm},t}\rho _{\mathtt{pm}}\alpha N_{\mathtt{pm}}^2\zeta_{\mathtt{pm},t}^3\mathbb{E}\big \{ \big |\hat{\mathbf{g} }_{m,1}^{T}\mathbf{G}_{m,\mathtt{pm}}^{*}\big |^2 \!\big \},\notag\\
   &= \sum\nolimits_{m\in\mathcal{M}_{c} }\eta _{\mathtt{pm},t}\rho _{\mathtt{pm}}\eta _{m,1}\rho _{\mathtt{c}}\zeta_{\mathtt{pm},t}\beta_{m,\mathtt{pm}}N_{\mathtt{pm}}^2N\gamma_{m,1}\notag\\
    &\times\big ( \beta_{\mathtt{pm},\mathtt{pm}}+\alpha N_{\mathtt{pm}}\zeta_{\mathtt{pm},t}^2 \big ).
\end{align}

6) Compute $\mathbb{E}\big \{ |\mathrm{SI}_{\mathtt{c}}|^{2} \big \} $:
\begin{align}
    &\mathbb{E}\big \{\! |\mathrm{SI}_{\mathtt{c}}|\!^{2} \big \}=\mathbb{E}\big \{ \big |\mathbf{w}_{\mathtt{comb},\mathtt{pm}}^T\sqrt{\eta _{\mathtt{pm},1}\rho _{\mathtt{pm}}}\mathbf{G}_{\mathtt{pm},\mathtt{pm}}^{T}\mathbf{g}_{\mathtt{pm},1} ^{\ast }\big |^2 \big \}\notag\\
    &+\mathbb{E}\big \{ \big |\mathbf{w}_{\mathtt{comb},\mathtt{pm}}^T\sqrt{\eta _{\mathtt{pm},t}\rho _{\mathtt{pm}}}\sqrt{\alpha}\mathbf{h}_{\mathtt{pm},t}\mathbf{h}_{\mathtt{pm},t}^{T}\mathbf {g}_{\mathtt{pm},1}^{\ast}\big |^2 \big \},\notag\\
    &=\sum\nolimits_{m \in\mathcal{M}_c }\eta _{\mathtt{pm},1}\eta _{m,1}\rho _{\mathtt{c}}\rho _{\mathtt{pm}}\gamma_{m,1}N\beta_{\mathtt{pm},\mathtt{pm}}N_{\mathtt{pm}}^2\beta_{m,\mathtt{pm}}\beta_{\mathtt{pm},1}\notag\\
    &+\sum\nolimits_{m \in\mathcal{M}_c }\eta _{\mathtt{pm},1}\eta _{m,1}\rho_{\mathtt{pm}}\rho _{\mathtt{c}}\alpha N_{\mathtt{pm}}^2\beta_{\mathtt{pm},1}\zeta_{\mathtt{pm},t}^2\beta_{m,\mathtt{pm}}N\gamma_{m,1},\notag\\
    &=\sum\nolimits_{m \in\mathcal{M}_c }\eta _{\mathtt{pm},1}\eta _{m,1}\rho _{\mathtt{c}}\rho _{\mathtt{pm}}\gamma_{m,1}NN_{\mathtt{pm}}^2\beta_{m,\mathtt{pm}}\beta_{\mathtt{pm},1}\notag\\
&\times(\beta_{\mathtt{pm},\mathtt{pm}}+\alpha\zeta_{\mathtt{pm},t}^2).
\end{align}

7) Compute $\mathbb{E} \{ |\mathrm{n}_{\mathtt{pm}}|^{2}  \}$:
\begin{align}\label{eq:n_j}
        &\mathbb{E}\big \{ |\mathrm{n}_{\mathtt{pm}}|^{2} \big \}=\sum\nolimits_{m \in\mathcal{M}_c } \eta _{m,1}\rho _{\mathtt{c}}\mathbb{E}\big \{ \hat{\mathbf{g}}_{m,1}^{T}\mathbf{G}_{m,\mathtt{pm}}^{\ast}    \mathbf{G}_{m,\mathtt{pm}}^{T} \hat{\mathbf{g}}_{m,1}^{\ast} \big \}, \notag \\
        &=\sum\nolimits_{m\in\mathcal{M}_c }\eta _{m,1}\rho _{\mathtt{c}}NN_{\mathtt{pm}}\beta_{m,\mathtt{pm}}\gamma_{m,1}.    
\end{align}
\ifCLASSOPTIONcaptionsoff
  \newpage
\fi




\balance
\bibliographystyle{IEEEtran}
\bibliography{IEEEabrv,Bibliography}

\begin{thebibliography}{10}
\providecommand{\url}[1]{#1}
\csname url@samestyle\endcsname
\providecommand{\newblock}{\relax}
\providecommand{\bibinfo}[2]{#2}
\providecommand{\BIBentrySTDinterwordspacing}{\spaceskip=0pt\relax}
\providecommand{\BIBentryALTinterwordstretchfactor}{4}
\providecommand{\BIBentryALTinterwordspacing}{\spaceskip=\fontdimen2\font plus
\BIBentryALTinterwordstretchfactor\fontdimen3\font minus \fontdimen4\font\relax}
\providecommand{\BIBforeignlanguage}[2]{{%
\expandafter\ifx\csname l@#1\endcsname\relax
\typeout{** WARNING: IEEEtran.bst: No hyphenation pattern has been}%
\typeout{** loaded for the language `#1'. Using the pattern for}%
\typeout{** the default language instead.}%
\else
\language=\csname l@#1\endcsname
\fi
#2}}
\providecommand{\BIBdecl}{\relax}
\BIBdecl

\bibitem{ref16}
Z.~Behdad, Ã.~T. Demir, K.~W. Sung, E.~Björnson, and C.~Cavdar, ``Multi-static target detection and power allocation for integrated sensing and communication in cell-free massive {MIMO},'' \emph{{IEEE} Trans. Wireless Commun.}, vol.~23, no.~9, pp. 11\,580--11\,596, Sept. 2024.

\bibitem{ref25}
B.~Liao, H.~Q. Ngo, M.~Matthaiou, and P.~J. Smith, ``Power allocation for massive {MIMO-ISAC} systems,'' \emph{{IEEE} Trans. Wireless Commun.}, pp. 1--1, 2024.

\bibitem{ref9}
M.~Elfiatoure, M.~Mohammadi, H.~Q. Ngo, and M.~Matthaiou, ``Cell-free massive {MIMO} for {ISAC}: Access point operation mode selection and power control,'' in \emph{Proc. IEEE GLOBECOM}, Dec. 2023, pp. 104--109.

\bibitem{ref:CYu_cellfree_ISAC}
Y.~Cao, Q.-Y. Yu, J.-C. Guo, and J.~Cheng, ``Voronoi-cluster multi-resolution hierarchical codebook design for cell-free integrated sensing and communication systems,'' \emph{{IEEE} Trans. Commun.}, vol.~72, no.~7, pp. 4432--4445, Jul. 2024.

\bibitem{10684238}
M.~Mohammadi, Z.~Mobini, H.~Q. Ngo, and M.~Matthaiou, ``Next-generation multiple access with cell-free massive {MIMO},'' \emph{Proc. {IEEE}}, pp. 1--49, 2024.

\bibitem{Liu:2022:survey}
A.~Liu \emph{et~al.}, ``A survey on fundamental limits of integrated sensing and communication,'' \emph{IEEE Commun. Surv. Tutor.}, vol.~24, no.~2, pp. 994--1034, Feb. 2022.

\bibitem{ref:yassen_TWC}
Y.~S. Atiya, Z.~Mobini, H.~Q. Ngo, and M.~Matthaiou, ``Secure transmission in cell-free massive mimo under active eavesdropping,'' \emph{{IEEE} Trans. Wireless Commun.}, pp. 1--1, 2024.

\bibitem{ref:zahra_iot}
Z.~Mobini, H.~Q. Ngo, M.~Matthaiou, and L.~Hanzo, ``Cell-free massive {MIMO} surveillance of multiple untrusted communication links,'' \emph{{IEEE} Internet Things J.}, 2024.

\bibitem{ref20}
C.~Zhong, X.~Jiang, F.~Qu, and Z.~Zhang, ``Multi-antenna wireless legitimate surveillance systems: Design and performance analysis,'' \emph{{IEEE} Trans. Wireless Commun.}, vol.~16, no.~7, pp. 4585--4599, May 2017.

\bibitem{ZAHRA:TIFS:2019}
Z.~Mobini, M.~Mohammadi, and C.~Tellambura, ``Wireless-powered full-duplex relay and friendly jamming for secure cooperative communications,'' \emph{{IEEE} Trans. Inf. Forensics Security}, vol.~14, no.~3, pp. 621--634, Mar. 2019.

\bibitem{ref18}
E.~Björnson, J.~Hoydis, and L.~Sanguinetti, ``Massive {MIMO} networks: Spectral, energy, and hardware efficiency,'' \emph{Foundations and Trends in Signal Processing}, vol.~11, no. 3-4, pp. 154--655, 2017.

\bibitem{ref12}
T.~M. Hoang, H.~Q. Ngo, T.~Q. Duong, H.~D. Tuan, and A.~Marshall, ``Cell-free massive {MIMO} networks: Optimal power control against active eavesdropping,'' \emph{{IEEE} Trans. Commun.}, vol.~66, no.~10, pp. 4724--4737, May 2018.

\bibitem{ref:perfect_CSI_at_monitor}
F.~Feizi, M.~Mohammadi, Z.~Mobini, and C.~Tellambura, ``Proactive eavesdropping via jamming in full-duplex multi-antenna systems: Beamforming design and antenna selection,'' \emph{{IEEE} Trans. Commun.}, vol.~68, no.~12, pp. 7563--7577, Dec. 2020.

\bibitem{ref15}
H.~Q. Ngo, A.~Ashikhmin, H.~Yang, E.~G. Larsson, and T.~L. Marzetta, ``Cell-free massive {MIMO} versus small cells,'' \emph{{IEEE} Trans. Wireless Commun.}, vol.~16, no.~3, pp. 1834--1850, Mar. 2017.

\bibitem{ref:jj_He_localization}
J.~He, Y.~J. Chun, and H.~C. So, ``A unified analytical framework for rss-based localization systems,'' \emph{{IEEE} Internet Things J.}, vol.~9, no.~9, pp. 6506--6519, May 2022.

\bibitem{ref14}
H.~Q. Ngo, L.-N. Tran, T.~Q. Duong, M.~Matthaiou, and E.~G. Larsson, ``On the total energy efficiency of cell-free massive {MIMO},'' \emph{IEEE Trans. Green Commun. Netw.}, vol.~2, no.~1, pp. 25--39, Mar. 2018.

\bibitem{ref4}
K.~Zhi \emph{et~al.}, ``Two-timescale design for reconfigurable intelligent surface-aided massive {MIMO} systems with imperfect {CSI},'' \emph{{IEEE} Trans. Inf. Theory}, vol.~69, no.~5, pp. 3001--3033, May 2023.

\end{thebibliography}

\vfill


\end{document}